# Self-suppression of the Giant CARS Background for Detection of Buried Interface with Sub-monolayer Sensitivity

CHANGHAO XU[1], YU ZHANG[1], QIANCHI FENG[1], RONGDA LIANG[1], CHUANSHAN TIAN[1,2,*]

[1]*State Key Laboratory of Surface Physics and Key Laboratory of Micro- and Nano-Photonic Structures (MOE), Department of Physics, Fudan University, Shanghai 200438, China*
[2]*Collaborative Innovation Center of Advanced Microstructures, Nanjing 210093, China*
*\*Corresponding author: cstian@fudan.edu.cn*

The past decades have witnessed marked progresses on the research of interfacial science in complex systems promoted by the advances in novel experimental techniques. Despite its success in many fields, implementation of coherent anti-Stokes Raman spectroscopy (CARS) for tackling the problems at interfaces was hindered by the huge resonant and non-resonant background from the bulk. Here we have developed a novel CARS scheme that is capable of probing a buried interface via suppression of the non-resonant and resonant bulk contribution by at least $10^5$ times. The method utilizes self-destructive interference between the forward and backward CARS generated in the bulk near the Brewster angle. As a result, we are able to resolve the vibrational spectrum of sub-monolayer interfacial species immersed in the surrounding media with huge CARS responses. We expect our approach not only opens up the opportunity for interrogation of the interfaces that involve apolar molecules, but also benefits other nonlinear optical spectroscopic techniques in promoting signal-to-background noise ratio.

## 1. INTRODUCTION

Interfaces are ubiquitous that unavoidably govern many important processes, e.g. superconductivity at metal oxide interface[1, 2], catalyst reaction[3, 4], electrochemistry[5, 6] and selective ion transport through biological membranes[7, 8]. Understanding of the properties and functionalities in these processes critically requires the knowledge of molecular and electronic structures at the microscopic scale. However, experimental probe of a buried interface that sandwiched between two media remains very challenging. In comparison to other surface-sensitive techniques[9-13], the second-order nonlinear optical spectroscopy, in particular sum-frequency generation (SFG) and second harmonic generation (SHG)[14-18], has been widely adopted for characterization of surfaces and interfaces in many fields and disciplines, because it offers *in situ*, remote and non-invasive probe of the interfacial species in the non-vacuum condition, for instance ambient or even liquid environment. Unfortunately, an intrinsic limitation of SFG/SHG is that it requires broken of inversion symmetry and is sensitive to polar molecules[16, 19-21] or those with large induced dipole at surface or interface[22-24], but is not suitable for the study of a large group of apolar species, such as hydrogen, oxygen, nitrogen and methane. While these species with centro-symmetry play pivotal roles in vast amount of energy and environmental related problems[25-28], it remains extremely difficult in experiment to unravel their molecular structure and reaction dynamics at the interfaces in practical environment, e.g. photocatalytic water splitting[29, 30] and fuel cells[31].

Coherent anti-Stocks Raman scattering (CARS) [32-35], being a third-order nonlinear wave mixing process, is a versatile label-free spectroscopic technique that can probe the Raman-active vibrational modes of molecules, including those with centro-symmetry. In principle, CARS with monolayer sensitivity may serve as a powerful analytical tool for *in situ* characterization of surfaces and interfaces[36-39]. However, because CARS is intrinsically not surface-specific, when applied to a buried interface, the CARS signal from the interface is easily overwhelmed by the giant bulk response as the pump laser beams passing through it. For example, a typical value of the coherent length ($L_c$) in the bulk is ~50 nm and ~10 μm for the backward and forward CARS geometry, respectively, which is about $10^2$-$10^4$ times larger than the thickness of an interface. Thus, to resolve the CARS spectrum of a buried interface, the dominating bulk resonant and non-resonant contributions must be suppressed by at least 4 orders of magnitude (if their third-order nonlinear susceptibilities are close to that at the interface), knowing the CARS intensity is proportional to the square of $L_c$. Several methods had been proposed to suppress the non-resonant background in CARS by, for instance, staggering time delay between excitation pulses and probe pulse[40, 41], arranging appropriate polarization combination in the detection scheme (polarization CARS)[42-44], destructive interference through shaping amplitude and phase of laser pulse [45] or interference with additional phase-locked CARS signal[46]. But, these approaches are not yet sufficient to suppress the bulk signal down to the level such that the vibrational spectrum from the buried interface is detectable, especially when the resonant background exists in the bulk.

In this paper, we present in theory and experiment a novel scheme that can detect the vibrational spectrum of a buried interface with sub-monolayer sensitivity after suppression of the resonant and non-resonant bulk contributions by at least $\sim 10^5$ times. Efficient self-suppression of the giant CARS response from the bulk is achieved when the phase matching angle is properly designed near the Brewster angle such that destructive interference occurs between the forward and the backward CARS. As a result, we are able to detect self-assembled monolayer of octadecyltrichlorosilane (OTS) and ethanol monolayer on a buried fused silica interface, respectively. The vibrational spectrum of those chemical groups with centro-symmetry that is absent in the SFG spectrum can now be clearly identified by the Brewster angle (BA)-CARS. We anticipate that our approach opens up the opportunity of interrogation of those buried interfaces where the physical or chemical microscopic processes involve apolar molecules, such as $CO_2$, $H_2$, $O_2$, $CH_4$ etc., that are abundant in electrochemistry and catalytical chemistry.

## 2. THEORY AND METHOD

As sketched in Fig. 2(a), we consider a buried interface between two nonlinear media with ultrathin interfacial layer. Being the third-order nonlinear optical process, CARS is not intrinsically surface specific. Thus, the CARS in reflection geometry consists of contributions from both the interface and the two neighboring bulk media[18]:

$$\vec{\chi}_{eff}^{(3)} = \vec{\chi}_{inter}^{(3)} + \int_{-\infty}^{0^-} \vec{\chi}_{B,I}^{(3)} \exp(i\Delta k_{z,I} z) dz$$
$$+ \int_{0^+}^{+\infty} \vec{\chi}_{B,II}^{(3)} \exp(i\Delta k_{z,II} z) dz \quad (1)$$

Here, $\vec{\chi}_{eff}^{(3)}$, $\vec{\chi}_{B,I}^{(3)}$ and $\vec{\chi}_{B,II}^{(3)}$ denotes the third-order nonlinear optical susceptibility from the interface, the bulk of Medium I and Medium II, respectively. $\Delta k_z = k_{sig,z} - k_{idl,z} + k_{800,z} - k_{CARS,z}$ is known as the phase mismatch, where $k_z$ is the projection of wavevector along surface normal (z-axis) and is defined positive for the forward propagation (+z direction) and negative for the backward propagation (-z direction). In the reflection geometry (backward propagation CARS, $k_{CARS,z} < 0$), the coherent length ($L_c = 1/\Delta k_z$) is about 50 nm, much larger than the thickness of the interface. In principle, the interfacial spectrum is detectable with resonant enhancement when the bulk contribution is non-resonant. Indeed, a monolayer of molecules on an exposed surface can be readily observed using CARS (solid curve in Fig. 2(c))[38].

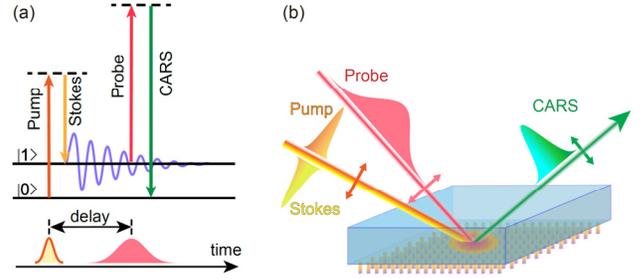

**Fig. 1.** Coherent Anti-Stokes Raman Scattering (CARS) for Probing a Buried Interface. (a), Diagram of CARS process. The Raman vibrations are excited by femtosecond broadband pump and Stokes pulses, followed by a picosecond probe beam after time delay (t). (b) Sketch of the experimental geometry for probing a buried interface in backward (BW) geometry. Polarizations of all four beams are set at P-polarizations, and the phase-matching angle of CARS is close to Brewster angle.

However, in the study of a buried interface, the situation becomes extremely difficult. Figure 2(c) presents the CARS spectrum of the bottom interface of silica with OTS deposited on (see sketch in the lower inset), where the interfacial OTS signal is totally overwhelmed by the non-resonant background by $10^5$ folds. The huge CARS contribution observed in experiment is missing in Eq. (1), which originates from the forward CARS in the bulk of medium I (purple ray in Fig. 2(a)). The forward CARS bears small phase mismatch and very long coherent length ($\sim 25\mu m$ in our experiment). It is several orders of magnitude stronger than the backward CARS. After reflection from the interface, it still dominates over all of the three contributions given in Eq. (1). Therefore, Eq. (1) shall be revised as the following:

$$\vec{\chi}_{tot,eff}^{(3)} = \vec{\chi}_{inter}^{(3)} + \int_{-\infty}^{0^-} \vec{\chi}_{B,I}^{(3)} \exp(i\Delta k_{z,I} z) dz$$
$$+ \int_{0^+}^{+\infty} \vec{\chi}_{B,II}^{(3)} \exp(i\Delta k_{z,II} z) dz + r\int_{-\infty}^{0^-} \vec{\chi}_{B,I}^{(3)} \exp(i\Delta k_{z,I}^{FW} z) dz$$
$$= \vec{\chi}_{inter}^{(3)} + \int_{-\infty}^{0^-} \vec{\chi}_{B,I}^{(3)} [\exp(i\Delta k_{z,I} z) + r\exp(i\Delta k_{z,I}^{FW} z)] dz$$
$$+ \int_{0^+}^{+\infty} \vec{\chi}_{B,II}^{(3)} \exp(i\Delta k_{z,II} z) dz \quad (2)$$

Here, $r$ is the reflection coefficient of the CARS E-field at the interface and $\Delta k_z^{FW} = k_{sig,z} - k_{idl,z} + k_{800,z} - k_{CARS,z}^{FW}$ describes the phase mismatch of the forward CARS with $k_{CARS,z}^{FW} > 0$. Integration of Eq. (2) yields:

$$\vec{\chi}_{eff}^{(3)} = \vec{\chi}_{inter}^{(3)} + \frac{i\vec{\chi}_{B,I}^{(3)}}{\Delta k_{z,I}^{FW}} (\frac{\Delta k_{z,I}^{FW}}{\Delta k_{z,I}} + r) + \frac{i\vec{\chi}_{B,II}^{(3)}}{\Delta k_{z,II}} \quad (3)$$

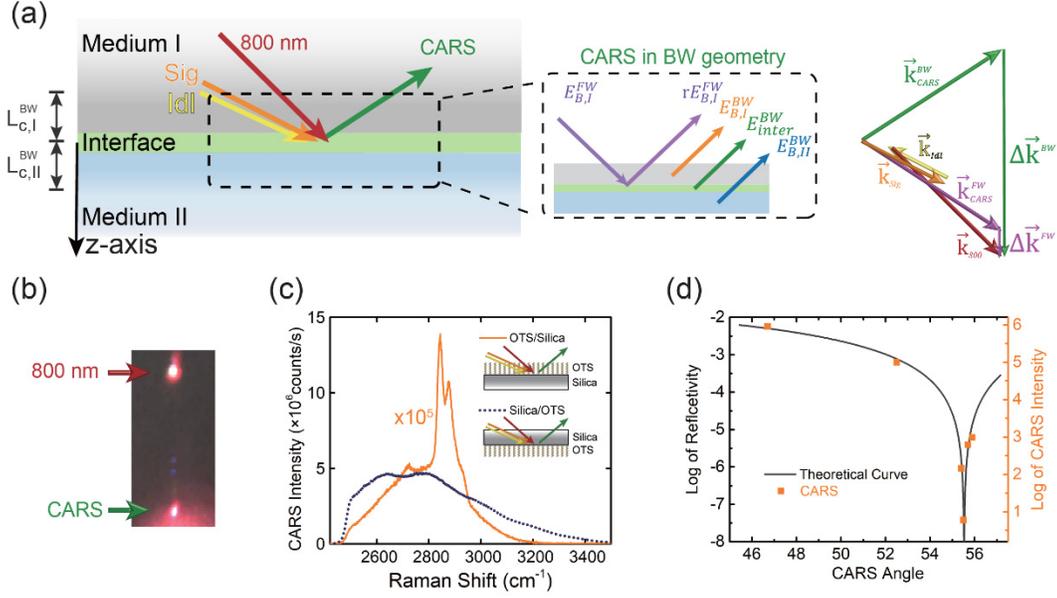

**Fig. 2.** Origins of CARS in probing of a buried interface and suppression of bulk responses using Brewster angle-CARS. (a) Sketch of CARS process near the buried interface between medium I and II excited by the signal (pump), the idler (Stokes) and the 800 nm (probe) pulses. The inset shows the four origins of CARS contributing in the backward (BW) geometry, namely reflection of forward (FW) CARS in Medium I ( $E_{B,I}^{FW}$ ) with long coherent length, BW CARS in the bulk of I ( $E_{B,I}^{FW}$ ) and II ( $E_{B,II}^{BW}$ ) with short coherent length $L_{c,I}^{BW}$ and $L_{c,II}^{BW}$, respectively, and CARS from the interface ( $E_{inter}^{BW}$ ). The drawing on the right illustrates the phase mismatch ( $\Delta \vec{k}$ ) in FW and BW CARS processes. (b) A photo for the beams of the 800 nm and FW CARS generated from fused silica in transmission, four-wave mixing between the signal and the idler. (c) The conventional CARS spectra of OTS on the top and bottom interface of silica as depicted in the inset. The spectrum for the former is enlarged by 105 times for comparison. (d) The calculated reflectance curve of P-polarized CARS from fused silica/air interface along with the measured CARS signal in BW geometry versus the phase matching angle.

Note, $\Delta k_{z,I} \sim \Delta k_{z,II} \gg \Delta k_{z,I}^{FW}$ and $\dfrac{\Delta k_{z,I}^{FW}}{\Delta k_{z,I}}$ is in the order of $10^{-2}$-$10^{-3}$. Thus, the second term dominates in Eq. (3). Fortunately, near Brewster angle, the reflection coefficient (*r*) of the P-polarized CARS can be tuned as small as the ratio of $\dfrac{\Delta k_{z,I}^{FW}}{\Delta k_{z,I}}$, and the sign of *r* is switchable when the angle is above or smaller than the Brewster angle. Therefore, the second term vanishes when the condition of $\dfrac{\Delta k_{z,I}^{FW}}{\Delta k_{z,I}} + r = 0$ is achieved. The obviously advantage of this scheme is that not only non-resonant background but the stronger resonant bulk signal in Medium I can be attenuated by many orders of magnitude with proper adjustment of *r*. Since the destructive interference is between the bulk signals their own (self-suppression), cancellation can be very efficient over broad spectral range.

To illustrate how strong the forward CARS is, Fig. 2(b) shows a picture of the non-resonant forward CARS in transmission generated in the bulk near the bottom interface of fused silica with polarization combination of PPPP and the CARS angle setting at 46.5° that is commonly used in experiment. The transmitted CARS signal is so strong that it is visually observable on a piece of paper. Using the same optical geometry but collecting CARS in the backward direction, the forward CARS after reflection is still ~$10^5$ times larger than that of monolayer OTS as compared in Fig. 2(c). In contrast, when the phase-matching angle approaches Brewster angle, the forward CARS decreases drastically. Figure 2(d) is the CARS intensity as a function of the CARS angle, which decreases from $9 \times 10^5$ at 46.5° to 6 at ~55.5°, in accordance with the calculated reflectivity curve. These results demonstrate that BA-CARS is an effective method to overcome the giant CARS signal from the bulk. More importantly, as will be shown in the following section, the above discussion is also valid for suppression of the resonant background.

## 3. EXPERIMENT

In our experiment, the tunable Raman excitation pulses and the probe beam were derived from a femtosecond (fs) mode-locked Ti:Sapphire laser and an associated optical parametric amplification (OPA) system. The fs signal (Sig, $\omega_{Sig}$) and the idler pulses (Idl, $\omega_{Idl}$) from OPA were aligned collinearly to excite Raman vibrations ($\omega_{vibs}$) with $\omega_{Sig} - \omega_{Idl} = \omega_{vibs}$. The picosecond (ps) probe pulse ($\omega_{800}$) was generated by passing the fs broadband beam centered at 800 nm through a home-build pulse shaper. The angle between the fs and the ps pulses was set at 10° in air. For the study of alkyl stretches, the center wavelengths of the Raman excitation and the probe pulses were chosen at 1280nm, 2140nm and 800 nm, with bandwidth of 50 nm, 155 nm and 1.2 nm, respectively. Accordingly, the center wavelengths for broadband CARS signal ( $\omega_{CARS} = \omega_{Sig} - \omega_{Idl} + \omega_{800}$ ) appears at 640 nm. The energy of the laser pulses was lowered down to the level such that no supercontinuum generation or damage was observed. CARS in the backward geometry is collected using EM-CCD after passing through a set of interference filters and monochromator. In our BA-CARS experiment, the polarization combination of PPPP were used in order to attenuate the reflection of all incoming beams, meanwhile the zzzz Raman tensor can be accessed that is sensitive to polar ordering at the interface. Optimization of beam overlapping was achieved using Au film or a 20 μm-thick polyethylene film coated on silica window.

## 4. RESULT AND DISCUSSION

**A. Probing OTS monolayer at buried interface by BA-CARS**

We have applied SFG, the conventional CARS and BA-CARS techniques to the study of OTS monolayer deposited on fused silica window as an example. Figure 3(a) shows the SFG spectrum of OTS/fused silica in the CH stretch region with S, S and P polarizations for the SF, the fixed visible (532 nm) and the tunable IR beams. The two strong bands at 2880 and 2940 cm$^{-1}$ were known as the symmetric stretch and the Fermi resonance of the terminal methyl group.[47] The absence of $CH_2$ mode at 2850 cm$^{-1}$ suggests the OTS molecules are in all-trans configuration, because the arrangement of methylene groups along a chain has inversion symmetry.[48] We then measure the OTS/fused silica sample with OTS on the top as illustrated in the inset of Fig. 3(b) using the conventional CARS scheme. Here, to suppress the non-resonant background, the probe pulse was delayed by 1.0 ps. The spectrum given in Fig. 3(b) clearly shows the features of CH stretches, which is complimentary to the SFG data. The strongest mode now is the symmetric stretches of $CH_2$ at 2850 cm$^{-1}$, accompanied by a weaker mode of the symmetric stretches of $CH_3$ at 2880 cm$^{-1}$. Considering there is only one $CH_3$ group and 17 $CH_2$ groups, the signal strength of the latter is obviously less than expected. It is because the methylene groups are nearly parallel to the surface and the projection of the P-polarized electric fields are small at the Brewster angle. Note that in probing OTS on the top surface, the phase matching angle of CARS is not critical. The angle for Fig. 3(b) was chosen at 46.5°, 9° away from the Brewster angle (55.5°). In this case, the non-resonant bulk signal strength with $L_c \sim$ 50 nm is comparable to that of those CH stretching bands.

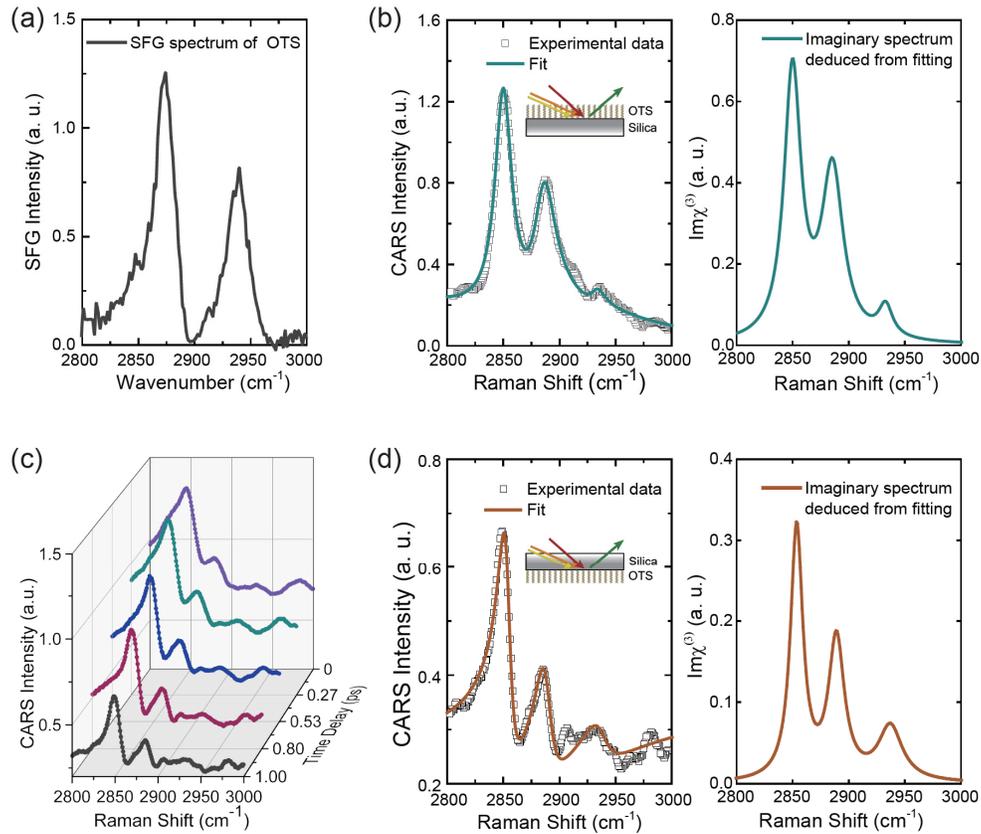

**Fig. 3.** OTS spectra probed by CARS and SFG spectroscopy. (a) SFG spectrum of OTS on silica. (b) CARS intensity spectrum of OTS on the top interface of silica with the probe pulse delayed by 1.0 ps. The spectrum was fit using three discrete modes, and the corresponding imaginary spectrum is presented on the right. (c) BA-CARS spectra of OTS on the bottom interface with staggering of time delay varying from 0 to 1.0 ps. (d) BA-CARS intensity spectrum taken from (c) for 1.0 ps delay time. The deduced imaginary spectrum is given on the right.

However, in the case of a buried interface, as has been demonstrated in Fig. 2(c), the conventional CARS does not work anymore, because it is now dominated by the huge background generated in the bulk of silica as the incident beams passing through it. By fine adjustment around Brewster angle, contribution of the forward CARS can be greatly diminished, while the interfacial response is less affected except for a little change in Fresnel factor. Figure 3(c) presents a set of BA-CARS spectra of OTS monolayer on the bottom surface of silica with the time delay of the probe pulse varying from 0 to 1.0 ps. Even without any time delay, the vibrational features of OTS are clearly observed since the background is suppressed down to the level comparable to the resonant bands of OTS. The resonant modes become more distinct as the non-resonant background is further suppressed by staggering the time delay. The imaginary spectra in Fig. 3(b) and 3(d), that were retrieved by fitting of the corresponding CARS intensity spectra assuming Lorentzian line-shape for discreate modes, are essentially the same, suggesting the success of BA-CARS for probing a buried interface.

### B. Probing monolayer of ethanol adsorbed on a buried interface by BA-CARS

To further verify the sensitivity of BA-CARS, we applied the technique to the study of a monolayer of molecules with low density of methylene and methyl groups. Ethanol ($CH_3CH_2OH$) adsorbed on fused silica surface was chosen for this purpose. It is known that ethanol forms a compact and ordered monolayer on silica as the ethanol partial pressure is above 2000 Pa.[49] We carried out BA-CARS measurement for the monolayer of ethanol on the bottom surface of silica (see Fig. 4(a)). The staggering time delay was set at 1.0 ps to further lower the non-resonant background. The measured BA-CARS spectrum and the imaginary component deduced from fitting are depicted in Fig. 4(b). Four resonant modes are observed at 2850 cm$^{-1}$, 2880 cm$^{-1}$, 2930 cm$^{-1}$ and 2970 cm$^{-1}$, assigned to the symmetric stretch of $CH_2$, the symmetric, Fermi resonance and antisymmetric stretches of $CH_3$, respectively. The comparable strength of the $CH_2$ and $CH_3$ bands indicates that the $CH_2$ group is tilted towards the surface normal, unlike the OTS case. [49] The results verify that BA-CARS is capable of probing a buried interface with sub-monolayer sensitivity via suppression of the giant non-resonant background.

### C. Probing a buried interface via suppression of the bulk resonant background

Besides the non-resonant background, the resonant contribution from the bulk is a more serious challenge in the study of interfaces. Similarly, the resonant bulk signal can be suppressed when $\frac{\Delta k_{z,I}^{FW}}{\Delta k_{z,I}} + r \sim 0$. To verify the validity of this method, we put the OTS/silica sample in a sealed chamber filled with high-pressure methane gas varying from 1 to 80 atmosphere (atm). The gaseous methane molecules contribute very large resonant CARS signal in the CH stretching region, even when the incident beams coming from the silica side (case I, as illustrated in Fig. 5(a)). Figure 5(b) presents the BA-CARS spectrum measured from the silica/OTS/methane interface (case I). As expected, the non-resonant background from silica is suppressed using BA-CARS. However, the OTS signal is hardly detectable because it is now overwhelmed by the large resonant spectrum of the high-pressure methane (65 atm). This can be understood by referring to Eq. (3). The bulk signal from the top medium (fused silica) is suppressed through destructive interference in the second term, but the third term originated from methane within the coherent length (~50 nm) is resonantly enhanced dominating the whole responses.

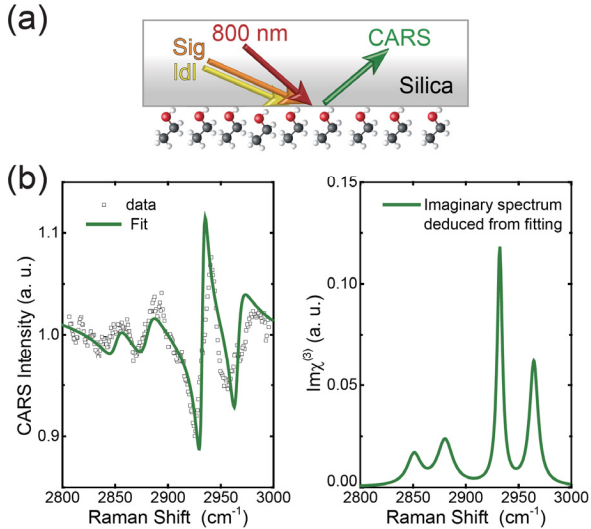

**Fig. 4.** Probing ethanol monolayer on buried interface with BA-CARS. (a) Sketch of the experimental arrangement for probing the silica buried interface. (b) BA-CARS Spectrum of ethanal monolayer on the buried interface. The solid curves represent the fitting results using four discrete modes.

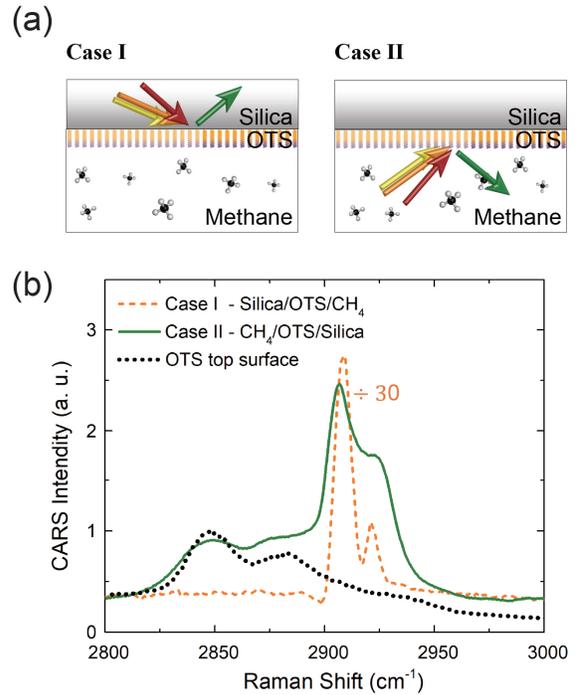

**Fig. 5.** Suppression of the resonant background in BA-CARS. (a) Sketch of the two experimental geometries used for the measurement of OTS/silica in high-pressure methane chamber. (b) Intensity spectra of methane and OTS obtained by BA-CARS in Case I and II. For comparison, the OTS CARS spectrum from Fig. 2(c) is reproduced here (dotted curve) in the region of 2800-3000 cm$^{-1}$.

To inhibit the strong resonant contribution, destructive interference of the methane response requires that the laser beams come from the opposite side of the interface, first going through the high-pressure methane gas (65 atm) (case II), and then collecting CARS in the backward geometry. The condition of $\frac{\Delta k_{z,1}^{FW}}{\Delta k_{z,1}} + r \sim 0$ can be achieved via adjusting the angle of the CARS beam around the Brewster angle, or alternatively fine tuning the Brewster angle by changing methane pressure. The result is presented in Fig. 5(b) (solid curve). The resonant signal of methane is attenuated by ~30 times smaller than case I, and the OTS spectrum can now be successfully resolved. The OTS bands at 2850 and 2880 cm$^{-1}$ agree well with that in Fig.2(c) (reproduced as dotted line in Fig. 5(b)), suggesting the monolayer OTS structure is robust against high pressure of gas. Note that the conventional CARS spectrum of methane is not compared here, since it is ~10$^6$ times stronger than the OTS in the optical geometry of case II. The residual of methane signal is due to limited extinction of polarization of the laser beams in our current experiment, which can be further suppressed with more accurate control of the laser polarization and extinction ratio. We therefore have verified that BA-CARS is also a valid scheme for probing a buried interface with the presence of bulk resonant background.

## 5. SUMMARY

We have demonstrated in this work a versatile and highly sensitive BA-CARS technique for probing buried interfaces. Besides those advantages of the coherent Raman spectroscopy, BA-CARS can be applied for the study of buried interfaces because not only the non-resonant background but more importantly the resonant responses from the bulk are suppressed by many orders of magnitude utilizing the destructive interference between the forward and backward CARS from the bulk near the Brewster angle. As a result, sub-monolayer of interfacial species can be resolved even when immersed in the surrounding media with huge CARS responses. Self-assembly monolayers of OTS and ethanol on the bottom interface of fused silica were studied as examples. In particular, the vibrational spectrum of methylene and methyl groups of the interfacial species were obtained in the presence of high-pressure methane gaseous environment. Clearly, the BA-CARS provides the crucial complimentary information to that of SFG for the understanding of an interface, because the former is sensitive to both polar and apolar species while the latter must require broken of inversion symmetry. Furthermore, with sub-monolayer sensitivity, BA-CARS offers the opportunity for in situ study of many crucial reactions at the interfaces immersed between two solid/liquid/gaseous media, including evolution of hydrogen, oxygen and methane at the interface of fuel cell[26], microscopic reaction pathways of $CO_2$ hydrogenation to CO, $CH_3OH$ and $CH_4$ [27, 28]etc. Note that the scheme presented in this work can also be applied to other nonlinear optical spectroscopic techniques when promotion of the signal-to-background noise is needed.

**Funding.** National Natural Science Foundation of China (No. 12125403 and No. 11874123); Shanghai Science and Technology Committee (No. 20ZR1406000).

**Disclosures.** The authors declare no conflicts of interest.

**Data availability.** Data underlying the results presented in this paper are not publicly available at this time but may be obtained from the authors upon reasonable request.